\documentclass[review]{elsarticle}

\usepackage{lineno,hyperref}
\usepackage{graphicx}
\usepackage{amsmath}
\usepackage{amsfonts}
\usepackage{amssymb}
\usepackage{url}
\usepackage{breakurl}
\usepackage[T1]{fontenc}
\usepackage{mathptmx}      
\modulolinenumbers[1]
\journal{Journal of Hydrology}

\begin{document}
\begin{frontmatter}
\title{Impact of large scale climate oscillation on drought in West Africa}


\author[mymainaddress]{Samuel, T. Ogunjo \corref{mycorrespondingauthor}}
\ead{stogunjo@futa.edu.ng, +2348026009902}
\author[mymainaddress,wascal]{Christiana, F. Olusegun}
\author[mymainaddress,mysecondaryaddress]{Ibiyinka, A. Fuwape}

\address[mymainaddress]{Department of Physics, Federal University of Technology, Akure, Ondo State, Nigeria}
\address[mysecondaryaddress]{Michael and Cecilia Ibru University, Ughelli North, Delta State, Nigeria.}
\address[wascal]{Department of Meteorology and Climate Science, West African Science Service Centre
on Climate Change and Adapted Land Use (WASCAL), FUTA}

\begin{abstract}
Drought poses a significant threat to the delicate economies in subsaharan Africa.  This study investigates the influence of large scale ocean oscillation on drought in West Africa.  Standardized Precipitation Index for the region was computed using monthly precipitation data from the Climate Research Unit during the period 1961 -1990.   The impact of three ocean oscillation indices - Southern Ocean Index (SOI), Pacific Decadal Oscillation (PDO) and North Atlantic Oscillation (NAO) on drought over West Africa was investigated using linear correlation, co-integration test, mutual information and nonlinear synchronization methods.  SOI showed predominantly positive correlation with drought over the region while PDO and NAO showed negative correlation.  This was confirmed by the co-integration tests.  The nonlinear test revealed more complex relationship between the indices and drought.  PDO has lesser influence or contribute less to the drought in the coastal region compared to the Sahel region of West Africa.
\end{abstract}

\end{frontmatter}
\newpage

\section{Introduction}
Variations in local and global climate are influenced by large scale (atmospheric and ocean) oscillations \citep{zhang2007}.  Pacific Decadal Oscillation (PDO), North Atlantic Oscillations (NAO) and El Nino Southern Oscillations (ENSO) are some of the large scale oscillations whose impacts are felt across different parts of Africa.   The North Atlantic Oscillation is a measure of the variations in lower atmospheric pressure over Iceland and higher pressure near the Azores and Iberian Peninsula \cite{JONES2014}.
NAO has been found to have impact on the hydroclimatic conditions in the Mediterranean region \citep{Brandimarte2011}, export of dust from North Africa \citep{moulin1997control}, precipitation in southeastern Africa \citep{maurice2001}, European airline traffic \citep{Rossello2011}, harmful algae blooms \citep{Baez2014} and population density of migratory birds \citep{anders2006}. Over West Africa, strong correlation has been found between Atlantic Nino Index 1 and drought indices based on CORDEX models \citep{Adeniyi2018}. According to \citet{anyamba2001ndvi}, South Atlantic SST and Global tropics SST show significant strong correlation with drought in Normalized Difference Vegetation Index (NDVI) over East Africa.  Positive correlation between ENSO and NDVI have also been reported for the have also been reported for the  African continent \citep{anyamba1996interannual}. Studies have established coherence between NAO and ENSO \citep{huang1998}, tropical SSTs  \citep{Rajagopalan} and East Asian summer monsoon-ENSO relationship \citep{wu2012}.  NAO has stronger influence over agricultural yields of selected crops in Spain than ENSO \citep{gimeno2002identification} as well as drought in different parts of Europe \citep{Bonaccorso2015}.

The Southern Oscillation Index is a large scale oscillation based on the sea level pressure between Tahiti and Darwin Islands.  The SOI is associated with sustained periods of negative (positive) episodes referred to as  El Nino (La Nina) and occurs every 2-7 years.  SOI, though originating in pacific, has been recognized as a major climatic factor that influences global weather events.  SOI has been shown to have influence on precipitation in Australia with a lag of 6 months \citep{mcbride1983}, inverse multiscale relationship with winter drought in China \citep{Liu2018}, precipitation in South Korea \citep{Jin2005}, rice production in India \citep{Bhuvaneswari2013}, wave climate and beach rotation \citep{Ranasinghe2004} and extreme temperature \citep{Zhong2017}.  PDO, which usually occurs every 20 - 30 years, is observed $20^o N$ of the North Pacific ocean.  PDO has dominant influence in the North Pacific with secondary signature in the tropics while the effect of SOI is predominantly in the tropics \citep{Mantua2002}.  It is related to, affected by but distinct from SOI \citep{HENLEY201742}. Relationship has been found between PDO and fish population in the Northwest Pacific \citep{Zhou2015}, global precipitation in the extratropics \citep{Shi2018} and climate variability in Mexico \citep{Park2017}.  It has been suggested that there exist a relationship between SOI and PDO \citep{Shi2017}.  Indirect influence of PDO on precipitation in continental United States through its influence on SOI has been proposed \citep{Shi2018}.  Studies on drought within West Africa include investigation into the impact of ENSO on precipitation in Nigeria \citep{Okeke2006}, impact of drought in the Volta Basin \citep{Oguntunde2017} and impact of Hardley Circulation on drought in Africa \citep{Nicholson}. Low-frequency signals in global Sea Surface Temperatures are now recognized as major drivers of the Sahelian drought \citep{Giannini1027}.

Large scale oscillations and climate variables have been shown to be either stochastic or chaotic \citep{ahn2005,Fuwape2017,Sivakumar2004,Fuwape2016b}.  The use of linear analysis in their investigation will lead to loss of information.  For example, \citet{Fuwape2016a} showed that the use of correlation overestimate the relationship between precipitation and radio refractivity in the tropics.  It is important to investigate complex and chaotic tendencies of climate time series for efficient short and long term predictions, model testing and better description \citep{Sivakumar2004}. Chaotic systems are sensitive to initial conditions and characterized by positive Lyapunov exponents.  Chaoticity and detrending of these systems tend to introduce spurious correlations and cross-correlations which will have significant impact on the results obtained in such studies and their interpretation \citep{Builes-Jaramillo2017,ogunjo2015effect}.  Recent development in chaos theory has provided a veritable tool to effectively study these nonlinear events without loss of information.
%
%

In this study, we aim to investigate the linear and nonlinear relationship between three large scale oscillation (PDO, NAO and SOI) on drought in the West African region. A  study of  linear analysis (correlation and cointegration tests) and dynamical tests (mutual information and phase synchronization) will not only give the performance of both tools but a robust relationship between large scale oscillations and droughts.

\section{Dataset and Methodology}

\subsection{Study area}

\begin{figure}
  \includegraphics[width=\textwidth]{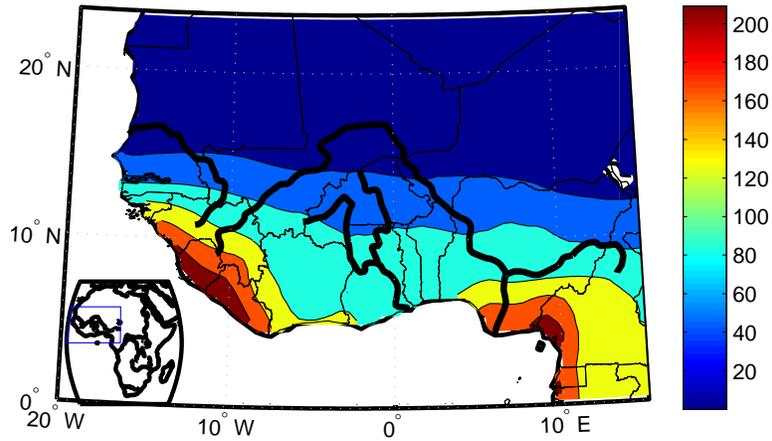}\\
  \caption{Map of study area showing the mean monthly precipitation (mm) during the period 1961- 1990.  The inset shows the study location within the continent of Africa.  Rivers are shown in thick black lines.}\label{fig1}
\end{figure}

The study area for this research is West Africa.  It lies within latitude $0^oN$ and $23.5^oN$ and longitudes $20^oW$ and $15^oE$ (Figure \ref{fig1}).  It is bounded in the south and west by the Atlantic Ocean and landed areas in the North and East.  The hydrographic network consist of short seasonal rivers around the Guinean Highlands, Gulf of Guinea, Niger Delta and Volta rivers.  \citet{omotosho1990} reported that deep convective system is responsible for 70\% of rainfall in the West African region.  The tropical diurnal convective systems that drives the West African Moonson involves the movement of moisture and heat between the Atlantic Ocean and continental land \citep{Vondou2010}.  Distribution of rainfall within the region is influenced by the position of the Intertropical discontinuity (ITD).  The ITD is separation between the tropical  maritime air mass and the dry tropical continental air mass.  It is a low pressure system which controls the spatial variation in tropical rainfall \citep{JosephO.Adejuwon2012}.

\subsection{Data}
\begin{figure}
  \includegraphics[width=\textwidth]{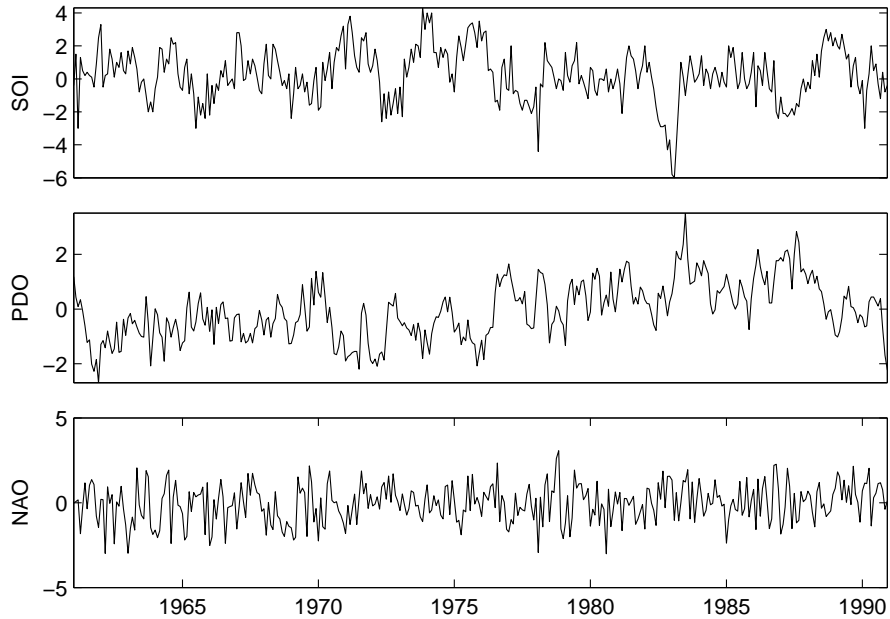}\\
  \caption{Monthly time series of climate indices  SOI (upper panel), PDO (middle panel), NAO (lower panel).}\label{fig2}
\end{figure}

Monthly time series for North Atlantic Oscillation (NAO) Index, Southern Oscillation (SOI) Index, and Pacific Decadal Oscillations (PDO) were obtained for 1961 - 1990 from The Earth System Research Laboratory, NOAA  \url{https://www.esrl.noaa.gov/psd/data/climateindices/list/}. Monthly precipitation data used was the version 4.01 of the Climate Research Unit Time Series (CRU TS - \url{http://data.ceda.ac.uk//badc/cru/data/cru_ts/cru_ts_4.01/data/pre/}) at a resolution of $0.5^o$ resolution from 1961 - 1990 \citep{harris2014}.  There were no missing data in both the climatic indices and precipitation data set. Anomalies of the climatic indices were used to remove annual variations. The temporal variation of the three climate indices under the period of study is shown in Figure \ref{fig2}.

\subsection{Standard Precipitation Index}
\begin{figure}
  \includegraphics[width=\textwidth]{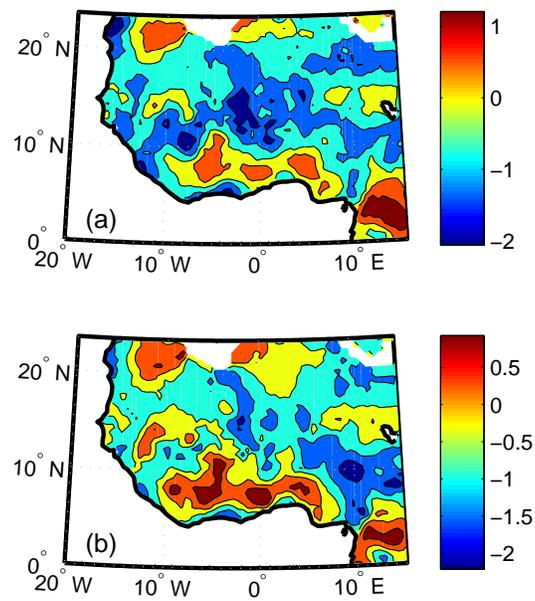}\\
  \caption{Spatial variation of SPI index over West Africa in (a) February 1965 and (b) June 1965 using a scale of 12 months.}\label{fig3}
\end{figure}

SPI was introduced by \citet{McKee1993179} to provide a drought indicator at different timescales based on probability.  In this approach, the precipitation data for a location is fitted to a gamma or Pearson Type III distribution before transforming to a normal distribution.  For precipitation values equal zero, the cumulative probability of the form
\begin{equation}\label{eq}
    H(x) = 1 + (1-q)G(x)
\end{equation}
where $q$ is the probability of zero precipitation \citep{Costa2011,Oloruntade2017}.  The values of SPI is interpreted as
\begin{equation*}
SPI = \left\{
  \begin{array}{ll}
    SPI \geq 2.00, & \hbox{Extremely wet;} \\
    1.50\leq SPI < 2.00, & \hbox{Very wet;} \\
    1.00\leq SPI < 1.50, & \hbox{Moderately wet;} \\
    -1.00\leq SPI < 1.00, & \hbox{Normal;} \\
    -1.50\leq SPI < -1.00, & \hbox{Moderately drought;} \\
    -2.00\leq SPI < -1.50, & \hbox{Severely drought;} \\
    SPI < -2.00, & \hbox{Extreme drought.}
  \end{array}
\right.
\end{equation*}
The use of SPI was recommend for drought monitoring by WMO \citep{svoboda2012standardized}.  SPI is easy to compute, can be calculated for different ranges of time (1, 3, 6, 13, 24 months), capture multitemporal  nature of rainfall deficiency and uses readily available precipitation data \citep{Shi2017}.  Typical values of SPI values during dry and wet seasons is shown in Figure \ref{fig3}.

\subsection{Linear correlation}
The interrelation between any climate index $x$ and SPI (represented by $y$) such that a change in one variable results in a positive or negative change in the other variable can be investigated using the correlation coefficient $\rho$.  The Pearson method of computing $\rho$ is described as
 \begin{equation}\label{pearson}
    \rho = \frac{\sum_{i=1}^n (x_i - \overline{x})(y_i - \overline{y})}{\sqrt{\sum_{i=1}^n (x_i - \overline{x})^2\sum_{i=1}^n (y_i - \overline{y})^2}}
 \end{equation}
The values of $\rho$ in the ranges $-1\leq\rho\leq0$ and $0\leq\rho\leq1$ represents positive and negative correlation respectively.  If $\rho = 0$, the two time series are said to be uncorrelated \citep{arora}.

\subsection{Co-integration}
Correlation (positive or negative) does not infer causality between two time series.  To determine causality between two time series the method of cointegration is used.  If the linear combination of two non-stationary time series, $x$ and $y$ are stationary after differencing, we say, the time series are cointegrated.  In this work, the Engle-Granger approach will be used \citep{engle1987co}.  The two time series are tested for stationarity, if not, they are made stationary by differencing with order 1 (ie I(1).  Ordinary least square is used to investigate the relationship $y_t = ax_t + e_t$.  The residual $e_t$ is checked for stationarity.  In this work, the linear relationship with constant trend is used.  Stationarity is tested using the Augumented Dickey-Fuller test.

\subsection{Mutual Information and sychronization}
Mutual information (I) quantifies linear and nonlinear interdependence between two systems and quantity of information shared \citep{Baptista2012}.  It is computed as
\begin{equation}\label{mutual}
    I(X;Y) = \sum_{x,y}P_{xy}(x,y) \log\frac{P_{xy(x,y)}}{P_x(x)P_y(y)}
\end{equation}
where $P_{x,y}$ and $P(x)$ are the joint and marginal distribution respectively.

An interdepence measure, H, was introduced by \citet{quiroga,quiroga1} which is easier to compute, robust to noise, and more sensitive to weak dependencies than mutual information.  It is expressed as
\begin{equation}\label{eqn_h}
    H^{(k)}(X|Y) = \frac{1}{N}\sum_{n=1}^N \ln \frac{R_n(X)}{R^{(k)}_n (X|Y)}
\end{equation}
$H$ is zero if X and Y are completely independent.  A new measure similar to $H$ but computed with a different averaging method is used in this study.
\begin{equation}\label{eqn_h}
    N^{(k)}(X|Y) = \frac{1}{N}\sum_{n=1}^N \ln \frac{R_n(X) - R^{(k)}_n (X|Y)}{R_n(X)}
\end{equation}
$N$ and $H$ have positive values if nearness in Y implies also nearness in X for equal time partner while negative values are obtained if close pairs in Y correspond to distant pairs in X \citep{quiroga,quiroga1}.

\section{Results and Discussion}

\begin{figure}
  \includegraphics[width=\textwidth]{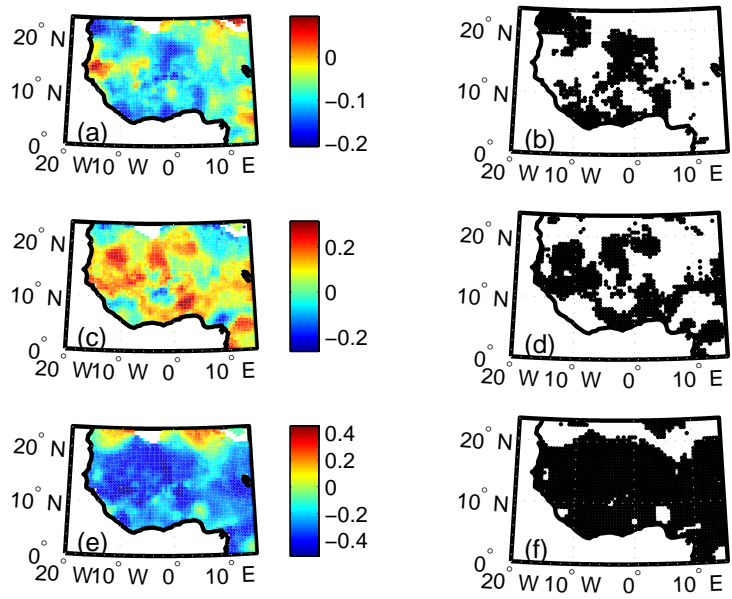}\\
  \caption{Zero lag correlation coefficient between SPI drought index and (a) NAO, (c) SOI and (e) PDO  and corresponding significance for (b) NAO, (d) SOI and (f) PDO} \label{fig4}
\end{figure}

Figure \ref{fig4} shows the correlation coefficient and its significance, between the teleconnection patterns considered and SPI drought for the West African region.  NAO shows a predominantly small but significant correlation with SPI drought index.  Positive but non-significant correlation could be observed along the Eastern coast of West Africa.  The Pacific Decadal Oscillation showed similar but stronger results over the same region (Figure \ref{fig4}e).  However, negative trends were not found along the Eastern Coast as in the case of NAO.   The Southern Oscillation Index showed the opposite trends observed in both NAO and PDO with Significant negative correlation distributed over the region.

\begin{figure}
  \includegraphics[width=\textwidth]{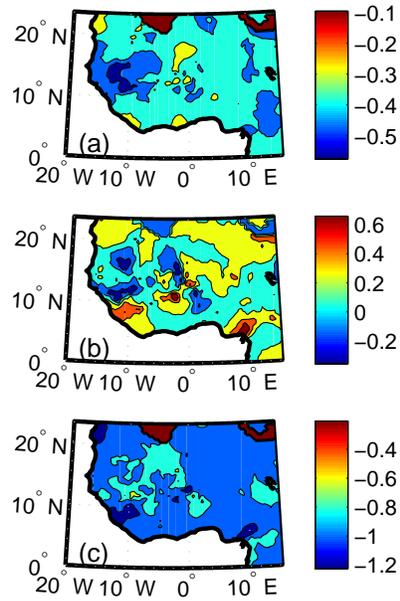}\\
  \caption{Engel-Granger co-integration test between SPI drought index and (a) NAO, (b) SOI and (c) PDO }\label{fig5}
\end{figure}

The result of the Engle-Granger cointegration test between large scale oscillations (PDO, NAO and SOI) and SPI index is shown in Figure \ref{fig5}.  Both NAO and PDO exhibit entirely negative linear relationship over the region. In the linear relationship between NAO and SPI, the eastern seaboard of West Africa showed the strongest negative correlation.  However, although SOI showed patches of negative relationship, the region boasts of predominantly positive linear relationship.

\begin{figure}
  \includegraphics[width=\textwidth]{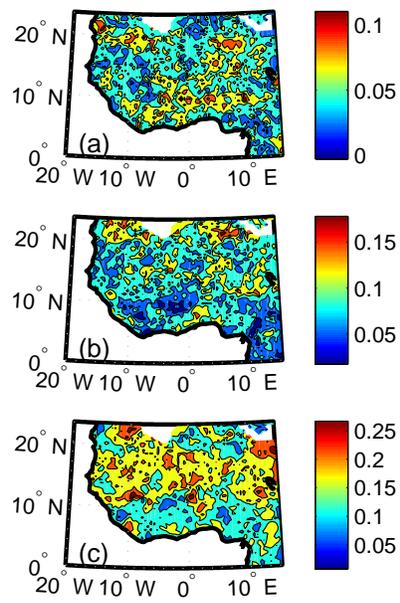}\\
  \caption{Mutual Information between SPI drought index and (a)  NAO, (b) SOI and (c) PDO }\label{fig6}
\end{figure}

The nonlinear relationships between the large scale oscillations (PDO, NAO, SOI) and SPI drought index over West Africa were investigated using mutual information and phase synchronization.  NAO showed weak mutual information with SPI in sparse locations (Figure \ref{fig6}).  This implies that not much information can be obtained concerning SPI by knowing the values of NAO. Similar patterns could be observed for SOI in the region. Large amount of information could, however, be obtained concerning SPI by having PDO values.  This is much more significant in the Sahel region of West Africa.

\begin{figure}
  \includegraphics[width=\textwidth]{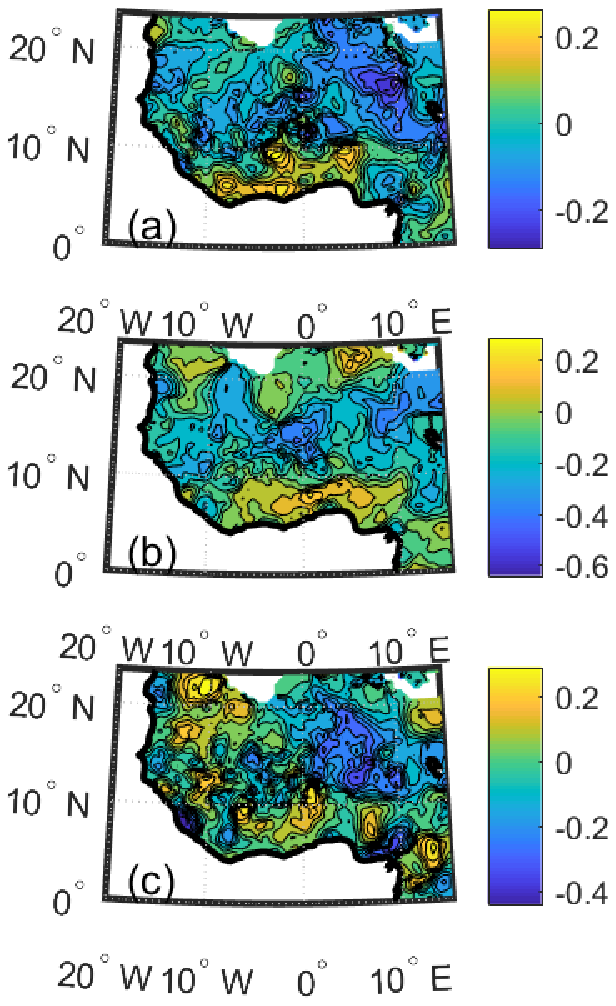}\\
  \caption{Nonlinear dependence measure, $N$, between SPI drought index and (a)  NAO, (b) SOI and (c) PDO }\label{fig8}
\end{figure}

Nonlinear dependence measure, $N$ was found to have predominantly negative values between NAO and SPI values (\ref{fig8}).  Positive values could be observed off the coast of Nigeria to Liberia which indicate nonlinear dependence.  SPI also showed positive nonlinear dependence with SOI along the coast of West Africa.  PDO showed patches of nonlinear dependence in the region.

Comparing the five different approaches to quantifying the relationship between drought in the West African region and large scale oscillation in both the Pacific and Atlantic Ocean, the mechanism of influence could be deduced.  It has been posited that sea surface temperature contribute significantly to drought regimes in West Africa but the specific contributions of major teleconnection patterns have not been investigated \citep{VISCHEL201995}.

Regions with highest negative correlation (dark blue) were found to correspond to regions with lowest co-integration (yellow).

\section{Conclusion}

Several factors such as land-atmosphere interaction, sea surface temperature and greenhouse gases  have been identified as drivers of drought in the West African region \citep{VISCHEL201995}.  These factors have spatio-temporal intensity and dominance for different parts of the region.  In this paper, the linear and nonlinear dependency between three large scale oscillation and drought index (SPI) in the West African region has been investigated.  This was achieved through the use of linear analysis (correlation coefficient) and nonlinear analysis (phase synchronization, nonlinear dependence test, and mutual information).

\end{document}